\documentclass[prd,aps,preprint,amsmath,nofootinbib,amssymb,eqsecnum,showkeys]{revtex4-1}
\pdfoutput=1
\usepackage{bm,verbatim,graphics,graphicx,color,slashed,textcomp}
\usepackage[colorlinks=true,
            linkcolor=red,
            urlcolor=blue,
            citecolor=blue]{hyperref}

\graphicspath{{figures/}}

\newcommand{\GeV}{\mathrm{GeV}}
\newcommand{\TeV}{\mathrm{TeV}}

\begin{document}

\title{Polynomial Spectrum of Gamma Ray from Dark Matter}
\author{Yong Tang}
\email{ytang@hep-th.phys.s.u-tokyo.ac.jp}
\affiliation{Department of Physics, Faculty of Science, \\
	The University of Tokyo, Bunkyo-ku, Tokyo 113-0033, Japan}

\begin{abstract}
In this study, we present some general features of gamma-ray spectra from dark matter. We find that the spectrum with sharp features could appear in a wide class of dark matter models and mimic the gamma line signals. If all other physical degrees of freedom are heavy or effectively decoupled, the resulting gamma ray from dark matter decay or annihilation would generally have polynomial-type spectra or power-law with positive index. We illustrate our findings in a model-independent framework with generic kinematic analysis. Similar results can also apply for cosmic ray or neutrino cases. 
\end{abstract}
\maketitle

\section{Introduction}
There are various compelling evidence for the existence of dark matter (DM) from sub-galactic length to cosmological scale. Searching for DM is therefore one of the primary tasks in current and future astro-particle physics experiments. Different experiments are designed for and sensitive only to some specific parameter space of enormous DM candidates~\cite{Jungman:1995df, Bertone:2004pz, Feng:2010gw}. Indirect detection of the annihilation or decaying products from DM can provide one of the most powerful complementary searches~\cite{Bergstrom:2000pn, Hisano:2004ds, Ishiwata:2008cu}. 

Gamma-ray experiments, such as Fermi-LAT~\cite{Atwood:2009ez}, can detect these energetic photons from DM self-annihilation or decay. The signature would show as an excess over the featureless, continuously falling background spectrum. These searches are much more sensitive to gamma rays with localized or sharp spectrum than with wide-spreading continuum spectra~\cite{Bringmann:2012ez}. It is widely believed that detection of a sharp gamma-ray line would be the smoking-gun for DM since, for instance, processes like DM+DM$\rightarrow 2\gamma$ can provide such line signals~\cite{Bergstrom:1997fj}.  

So far, only a few cases in which sharp gamma-ray spectra other than lines have been found, such as internal bremmstrahlung~\cite{Bringmann:2007nk, Bringmann:2012vr} and box-shaped signals~\cite{Ibarra:2012dw, Chu:2012qy, Ibarra:2013eda}, see Ref.~\cite{Bringmann:2012ez} for a recent review. However, to generate such kinds of sharp spectra, the underlying particle physics theories for DM usually have to satisfy some specific requirements. For example, to have internal bremmstrahlung the mediating particle needs to be electroweak charged~\cite{Toma:2013bka, Giacchino:2015hvk}, while box-shaped signals require the final on-shell particles have masses close to DM and decay sequentially into two photons. Other shapes could arise in more complicated models~\cite{Kim:2015usa}. To test such kind of theories, phenomenological and experimental analysis need to be performed model by model. 

In this paper, we show in a wide class of dark matter models there exist gamma rays with sharp spectra that can mimic the line signals, especially for heavy dark matter when the resulting gamma-ray's energy can not be resolved well enough. The generic feature of the spectrum is that the differential gamma-ray flux is a polynomial function of the energy. Our discussions are based on model-independent kinematics analysis, which provide a very efficient phenomenological framework for various DM theories or effective interactions. The presented method can also be used for searching neutrinos and cosmic rays from DM. 

This paper is organized as follows. In Sec.~\ref{sec:framework}, we establish the general theoretical framework for investigating the generated spectrum from DM decay and annihilation. Later in Sec.~\ref{sec:gammaray} we illustrate how to use the formalism to calculate gamma-ray spectrum and give the general basic polynomial functions. In Sec.~\ref{sec:application}, we show how polynomial/power-law spectrum can give rise to sharp spectra shape and mimic line signals. Finally, we summarize and conclude. 

\section{General Theoretical Framework}\label{sec:framework}

Theoretically, there are enormous particle physics models for DM with mass ranging from sub-eV to Planck scale due to our limiting knowledge of DM particle identities, see Refs.~\cite{Bertone:2004pz, Feng:2010gw} for recent reviews. For example, one popular DM candidate, weakly interacting massive particle (WIMP), has mass around $\mathcal{O}(10\GeV)$ to $\mathcal{O}(10\TeV)$ and is very attractive since its self-annihilation may give possible signatures for indirect detection. Another candidate, DM from inflation or reheating dynamics, could have mass as heavy as $10^{14}\GeV$~\cite{Chung:1998ua}. Therefore, in this study, other than focusing on a specific particle physics model, we instead consider some model-independent features. We limit our discussions to heavy DM and assume all new or mediating particles are heavy, compared to standard model ones, which can lead us to an efficient and model-independent way for phenomenological studies. 

In the framework of effective field theory, our discussions may begin with the following effective operators after integrating the heavy particles,
\begin{equation}
\delta \mathcal{L}=\sum_{i,j}\frac{\alpha_{ij}}{\Lambda^{d_{ij}-4}} \mathcal{O}^i_{X}\mathcal{O}^j_{\textrm{SM}},
\end{equation}
where $\mathcal{O}^{j}_{\textrm{SM}}$ are composite operators of standard model fields,  $\mathcal{O}^{i}_{X}$ can be a single field or composite operators of dark sector fields, $d_{ij}$ is the mass dimension of $\mathcal{O}^i_{X}\mathcal{O}^j_{\textrm{SM}}$, and  $\Lambda$ is the effective mass scale with corresponding coupling constant, $\alpha_{ij}$.  From the effective theory's perspective, when focusing the gamma-ray spectrum, one can impose either $SU(3)_C\times SU(2)_L \times U(1)_Y$ symmetry on $\mathcal{O}^{j}_{\textrm{SM}}$, or just the unbroken $SU(3)_C\times U(1)_Q$ symmetry instead. In the later case, photon field, $A_{\mu}$, may be of primary importance. Since we are interesting in the generic features, we shall not limit our later discussions to any specific operators. 

\begin{figure}[tb]
	\centering
	\includegraphics[scale=0.75]{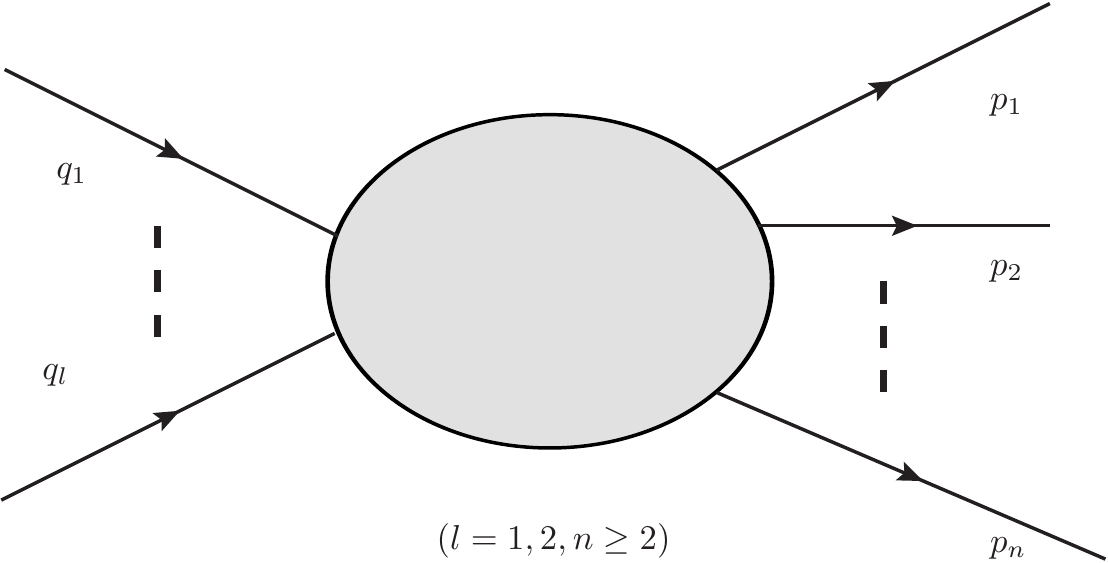}
	\caption{Feynman diagram for general $l$-to-$n$ process. $l=1$ corresponds to the usual decay process and $l=2$ for two-particle annihilation. $q_i$s and $p_f$s are momenta for initial and final particles. $q_i$s can be treated as non-relativistic $(m_i,0,0,0)$. \label{fig:feynman}}
\end{figure}

To investigate the spectrum of generated particles from DM decay or annihilation, a systematic way to view the general $l$-to-$n$ process can be illustrated in terms of Feynman diagram shown in Fig.~\ref{fig:feynman}. Although $l=1$ and $l=2$ correspond to the most interesting and widely studied decay and annihilation processes, respectively,  here we keep an open mind for general $l$ in cases where special types of interactions dominate. We assume all final states are standard model particles. Since from astrophysical observation, we have already known that DM particles are moving non-relativistically, we can replace initial momenta $q_i$ with $(m_i,0,0,0)$. In general $m_i$s do not have to be the same since some models could have multiple DM components. However, for our purpose here, only the total mass $M=m_1+\cdots+m_l$ is relevant.

The probability function $\sigma_l$ or normalized distribution for a final particle over its phase space is generally given by, for example, for particle $\#1$ with $p_1=(E_1,\bm{p}_1)$,
\begin{equation}\label{eq:master}
\frac{d\sigma_l}{\sigma_l d^3\Omega_1}=\frac{4\pi^2}{E_1 }\frac{d\sigma_l}{\sigma_l dE_1} = 
\dfrac{\displaystyle\int d^3\Omega_2...d^3\Omega_n \delta^3\left(\bm{p}_1+\sum_{f=2}^{n} \bm{p}_f\right)\delta\left(E_1 + \sum_{f=2}^{n} E_f-\sqrt{s}\right)\left|\mathcal{M}\right|^2}
{\displaystyle\int d^3\Omega_1...d^3\Omega_n \delta^3\left(\sum_{f=1}^{n} \bm{p}_f\right)\delta\left(\sum_{f=1}^{n} E_f-\sqrt{s}\right)\left|\mathcal{M}\right|^2},
\end{equation}
where $s=q^2\equiv\left(q_1+q_2+\cdots+q_l\right)^2\simeq M^2$, the phase space element has the following form,
\[
d^3\Omega\equiv \dfrac{d^3\bm{p}}{\left(2\pi\right)^3 2E},\;\bm{p}=(p_x,p_y,p_z),
\]
and $\left|\mathcal{M}\right|^2$ is the polarization-summed and squared matrix element. For signals that travel directly to detectors, like gamma ray or neutrino, we can relate them to the observable quantity, differential flux, through
\begin{equation}\label{eq:signal}
\frac{d \Phi}{d E_1} = \frac{1}{4\pi}\frac{d\sigma_l}{dE_1}{\displaystyle \int  dr}\left(\frac{\rho_{\textrm{DM}}(r)}{m}\right)^{l},
\end{equation}
where $r$ is the distance from observation point to decay or annihilation point and the integration is performed along the line-of-sight. 

In the above formula, Eq.~\ref{eq:signal}, we have kept the general $l$ for the initial states, although only the cases $l=1$ and $l=2$ are more important. By dimensional analysis and neglecting the numerical factor, the ratio of contributions from $l$ to $(l-1)$ is $\sim \rho_{\textrm{DM}}/m^4$ which is a dimensionless quantity and can be easily checked and verified with the usual decay and annihilation processes. $\rho_{\textrm{DM}}/m^4$ is usually very small, $10^{-39}$ for typical $\rho_{\textrm{DM}}\sim \GeV/\textrm{cm}^3$ and $m \sim \GeV$. So normally we can ignore high-$l$'s contributions if low-$l$ processes are not forbidden by kinematics or symmetry. However if the DM density is as high as that in neutron star $\rho\sim 10^{38}\GeV/\textrm{cm}^3$, we might need to consider high-$l$'s contributions. 

We also should note that going to multiple final states is not always sub-dominant because of phase space suppressing factor, $1/4\pi^2$. Actually, for heavy dark matter in some models, see Ref.~\cite{Ko:2015nma} for example, when processes with one more final state are considered, they are accompanied by $\dfrac{1}{4\pi^2}\dfrac{m^2_X}{v^2}$, where $m_X$ is the DM mass and $v$ is the electroweak breaking parameter $v\simeq 246$GeV. An easy check is that when $m_X>\TeV$ processes with multiple final states are more important. This is just one explicit case in which processes with mutliple final states could be the dominant contributions. 

The squared matrix element $\left|\mathcal{M}\right|^2$ is determined by the underlying particle physics theories, or effective interaction operators. It can have various, complicated forms, as a function of all Lorentz-invariant $p_i\cdot p_j$, 
\begin{equation}
\left|\mathcal{M}\right|^2 = f(p_i\cdot p_j),
\end{equation}
where $p_i$, without confusion here, stands for both initial and final momenta, $q_i$ and $p_f$. The above formula used only Lorentz invariance of $\mathcal{M}$. Furthermore, if all unknown or new particles that appear virtually in the bubble of Fig.~\ref{fig:feynman} are heavy, much heavier than $m_i$, we can reduce $\left|\mathcal{M}\right|^2$ to a general polynomial function of momenta,
\begin{equation}
\left|\mathcal{M}\right|^2 = C_0 + C_{ij}\ p_i\cdot p_j + C_{ijkl}\ p_i\cdot p_j\ p_k\cdot p_l + \textrm{higher powers of $p_i$},
\end{equation}
where all the coefficients $C_0, C_{ij}$ and $C_{ijkl}$ are constants and can be calculated from the explicit effective operators.


\section{Features of Gamma-Ray Spectrum}\label{sec:gammaray}

The above formalism are very generic and can be used to calculate the production spectrum of any particle in the final states, such as photon, neutrino, positron/electron and proton/anti-proton. One of the particularly interested messengers is gamma ray since it can travel without deflection and point to the source. For the SM final states, without loss of generality and to get analytic compact results, we can simply assume they are all massless. This approximation is well justified when we consider heavy DM with mass larger than $\mathcal{O}(\TeV)$, which is the region of particular interest to future ground experiments.  

Let us start with very simple and familiar cases. From eq.~\ref{eq:master}, we can immediately infer that for two-body final states, the resulting distribution is always mono-energetic, namely a $\delta$-function, $\delta\left(E_1-M/2\right)$. This is also valid for processes with any number of initial or incoming states. Interesting particle physics models that give such kind of gamma-line signatures include, DM decay, DM$\rightarrow \nu+ \gamma$, annihilation into two photons, DM+DM$\rightarrow \gamma+\gamma/Z$ and so on. 

For three-body final states, it would be much more complicated due to various possible $\left|\mathcal{M}\right|^2$. Again let us first consider the simplest that $\left|\mathcal{M}\right|^2=C_0$ which is constant. One operator that can give a constant $\left|\mathcal{M}\right|^2$ is $\mathcal{O}_{X}hA_\mu A^\mu$ which could result from a particle physics model where there exists kinetic mixing between extra $U(1)$ and the $U(1)$ symmetry in standard model.  After performing the integral in numerator, we get the {\it distribution function},
\begin{equation}\label{eq:case1}
\frac{dP_l}{dx}\equiv\frac{d\sigma_l}{\sigma_l dE_1} = 8x,\ 0\leq x\equiv E_1/M \leq 1/2,
\end{equation}
which is a linear function of $E_1\equiv xM$ and $1/2$ is the kinematics endpoint. A slightly complicated situation is that $\left|\mathcal{M}\right|^2= q_i\cdot p_1$ when $F_{\mu\nu}=\partial_\mu A_\nu-\partial_\nu A_\mu$ appears in $\mathcal{O}_{\textrm{SM}}$, and we can obtain $dP_l/ dx = 24x^2$ straightforwardly. Those two examples show that the resulting gamma ray can have power-law spectrum with a positive index.

For $\left|\mathcal{M}\right|^2\propto q_i\cdot p_j(j\neq 1)$, then we get 
\begin{equation}\label{eq:case2}
\frac{dP_l}{dx} = 12x(1-x),\ 0\leq x\leq 1/2.
\end{equation}
all other possibilities with bilinear $p_i\cdot p_j$ can be reduced to the above three bases. For instance, $p_2\cdot p_3=\left(M^2-2q\cdot p_1\right)/2$ and $p_1\cdot p_2=\left(M^2-2q\cdot p_3\right)/2$. 

We can continue to investigate cases with higher powers of momenta for more complicated effective interactions. For instance, fermionic fields or derivatives of final states would contribute more $p$s in $\left|\mathcal{M}\right|^2$, so $\mathcal{O}_{\textrm{SM}}=\bar{\psi}\sigma_{\mu\nu}\psi F^{\mu\nu}$ would lead to terms like $p_i\cdot p_j\ p_k\cdot p_l$, see Refs.~\cite{Ko:2015nma, Aisati:2015ova} for other concrete examples. However, it is easy to convince oneself that the general formula would be polynomial functions of $E_1$ or $x$,
\begin{equation}\label{eq:xpoly}
\frac{dP_l}{dx} = D_1\times 8x + D_2\times 24x^2 + D_3\times 64x^3 +\cdots =\sum_{i=1} D_i\left(i+1\right)2^{i+1}x^i,\ 0\leq x\leq 1/2,
\end{equation}
where $D_i$ are dimensionless constants with $\displaystyle \int dx \dfrac{dP_l}{ dx} =1$ or $\sum_i D_i=1$, but their precise values are determined by the underlying complete theories or effective operators. One-to-one correspondence between $D_i$ and all standard model effective operators would require other dedicated analysis, which is beyond our scope here. 

We can check that Eqs.~\ref{eq:case1} and \ref{eq:case2} respectively correspond to
\[
D_1=1, D_{i\neq 1}=0 \;\ \textrm{and }\ \; D_1=\frac{3}{2}, D_{2}=-\frac{1}{2}, D_{i> 2}=0. 
\]

The above result, eq.~\ref{eq:xpoly}, is also true for cases with more than three final states, $n>3$. This can be proved by mathematical induction, or we can just perform the phase space integration in the numerator of Eq.~\ref{eq:master} which would be a polynomial function of $m^2_{23\cdots n}\equiv\left(p_2+p_3+\cdots + p_n\right)^2=M^2-2ME_1$. Explicit expansion of a polynomial function of $m^2_{23\cdots n}$ then gives polynomial on $E_1$. 

One thing we should point out is that the polynomial form is valid in the massless approximation for final particles and under this approximation, we can get compact analytical form. In case of massive final states, we should anticipate there are also terms like $x^i (\ln x)^j$ which are sub-dominant and can be ignored when DM mass is much heavier that SM particles. 

So far we have only concentrated on the ``primary" photons which are produced directly from DM decay or annihilation. There are also ``secondary" photons which result from the electromagnetic cascade from other final states, such as leptons, quarks, gauge bosons and Higgs particle. Such photons are usually subdominant and much softer, therefore would not affect the shape features at the high energy part. Nevertheless, they can be calculated by convolution,
\begin{equation}
\frac{dP^{\textrm{sec}}_l}{dx}=\int dx'\frac{dP^{\textrm{pri}}_l}{dx'}\frac{dN(x')}{dx},
\end{equation}
where $dP^{\textrm{pri}}_l/dx'$ in the integrand is calculated just like in previous discussion, and $dN(x')/dx$ is the number distribution for primary particle with energy $x'M$ giving photons with energy $xM$. $dN/dx$ can be obtained by using standard events generator, such as {\texttt{\textbf{pythia}}}~\cite{Pythia64:2006za}.

\section{Applications and Comparisons}\label{sec:application}
\begin{figure}
\includegraphics[width=0.48\textwidth,height=0.5\textwidth]{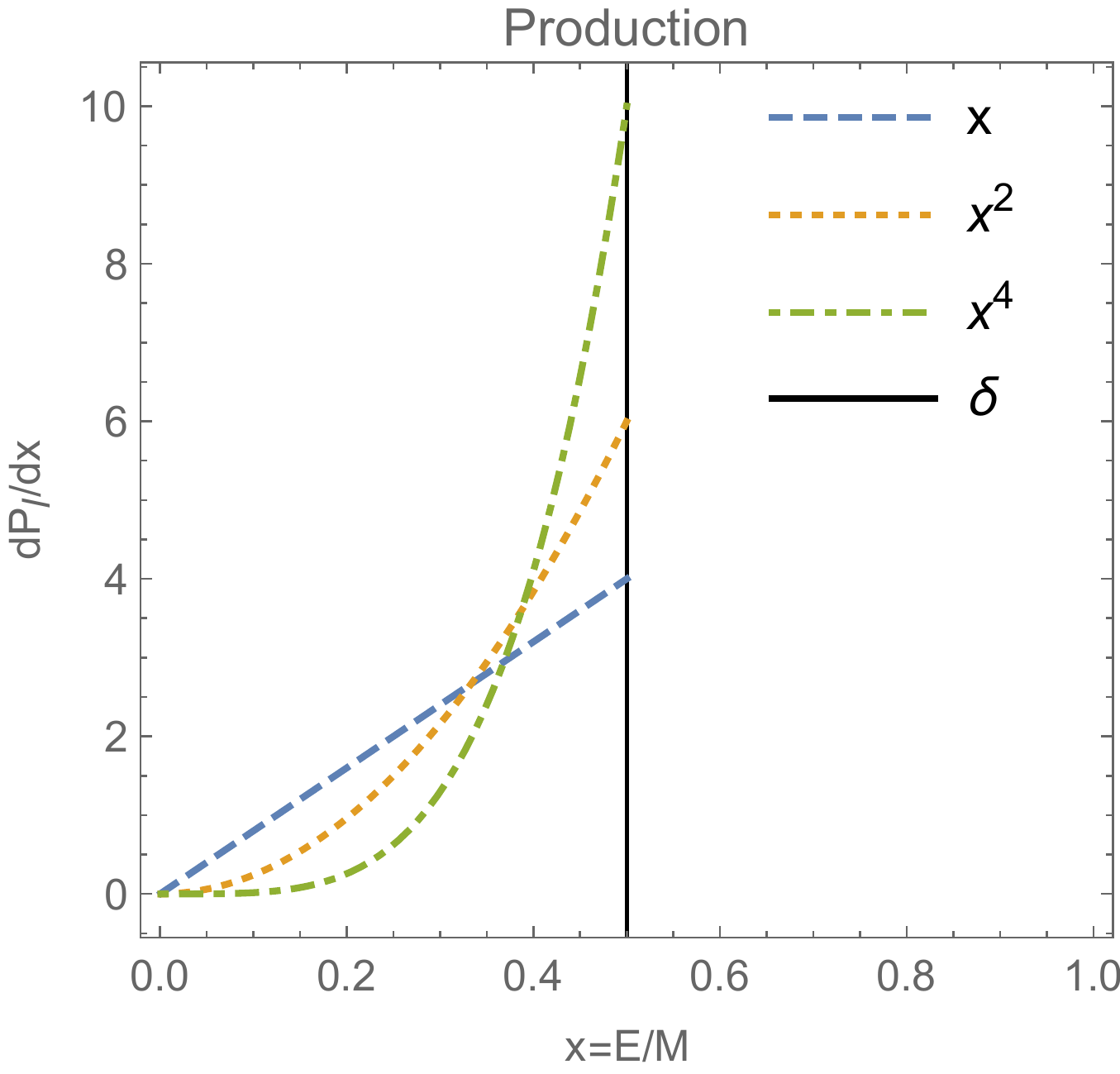}
\includegraphics[width=0.48\textwidth,height=0.5\textwidth]{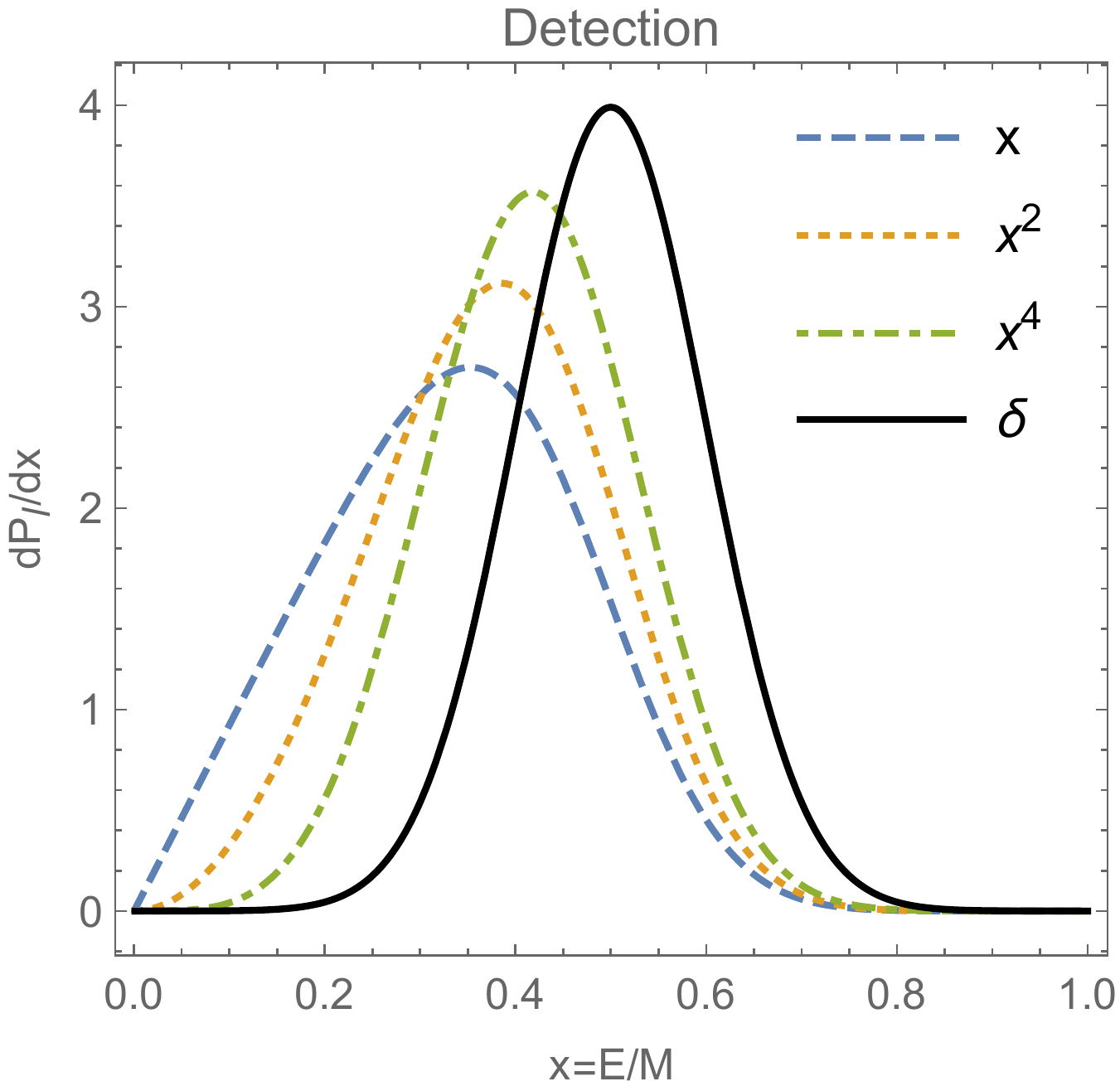}
\caption{Energy distributions at production (left panel) and detection (right panel) for different spectrum. The energy cut at $x=1/2$ is due to kinematical endpoint. All spectra are normalized. See text for details. \label{fig:spectral}}
\end{figure}

In this section, we show how sharp spectra for gamma ray can arise from power-law/polynomial spectrum and mimic the standard gamma-ray line signals. For phenomenological studies, one can either start with an UV complete theory or an effective operator and calculate the gamma-ray spectrum, or just assume some single/mixed power law spectra without specifying its particle physics origin. We will take the latter approach in this section. 

Due to the finite energy resolution of gamma-ray detector, the spectra measured or reconstructed can not perfectly show the original features at production point. For example, a monochromatic line ($dP_l/dx=\delta(x-1/2)$) would display as a Gaussian distribution~\footnote{The real situation may be more complicated. For example, Fermi-LAT collaboration convoluted with three Gaussian functions~\cite{Ackermann:2015lka}.},
\begin{equation}\label{eq:convolution}
\frac{dP_l}{dx}=\int dx' \frac{\delta(x'-\frac{1}{2}) }{rx'\sqrt{2\pi}}\exp \left[-\frac{(x-x')^2}{2r^2x'^2}\right],
\end{equation} 
where $r$ is the energy resolution, typical at order of $0.2$ when energy is as high as $\mathcal{O}(\TeV)$. The behaviors can be seen from the black lines in Fig.~\ref{fig:spectral} where the left panel gives the spectra at production and the right gives the expected ones from detection.

Now we simulate spectrum with power law at production and convolute with Gaussian energy dispersion. We illustrate with three simple cases,
\begin{equation}
\frac{dP_l}{dx} = \left(i+1\right)2^{i+1} x^i,\ 0<x<1/2, \; i=1,2,4, 
\end{equation}
where the constant coefficients are due to normalization, Eq.~\ref{eq:xpoly}. Their shapes are shown in Fig.~\ref{fig:spectral} as dashed, dotted and dash-dotted curves, respectively. As shown, the reconstructed or expected spectra can mimic Gaussian-like signals with displaced central values and broader widths. The larger index power-law has, more similar and closer to Gaussian distribution (black curve). These findings suggest that if in future experiments a line-signal is detected, it may also be explained by or identified as power-law signals. Or equivalently, we can search for general polynomial-type signals with several parameters other than just gamma lines. 

\begin{figure}[tbh]
	\includegraphics[width=0.4\textwidth,height=0.3\textwidth]{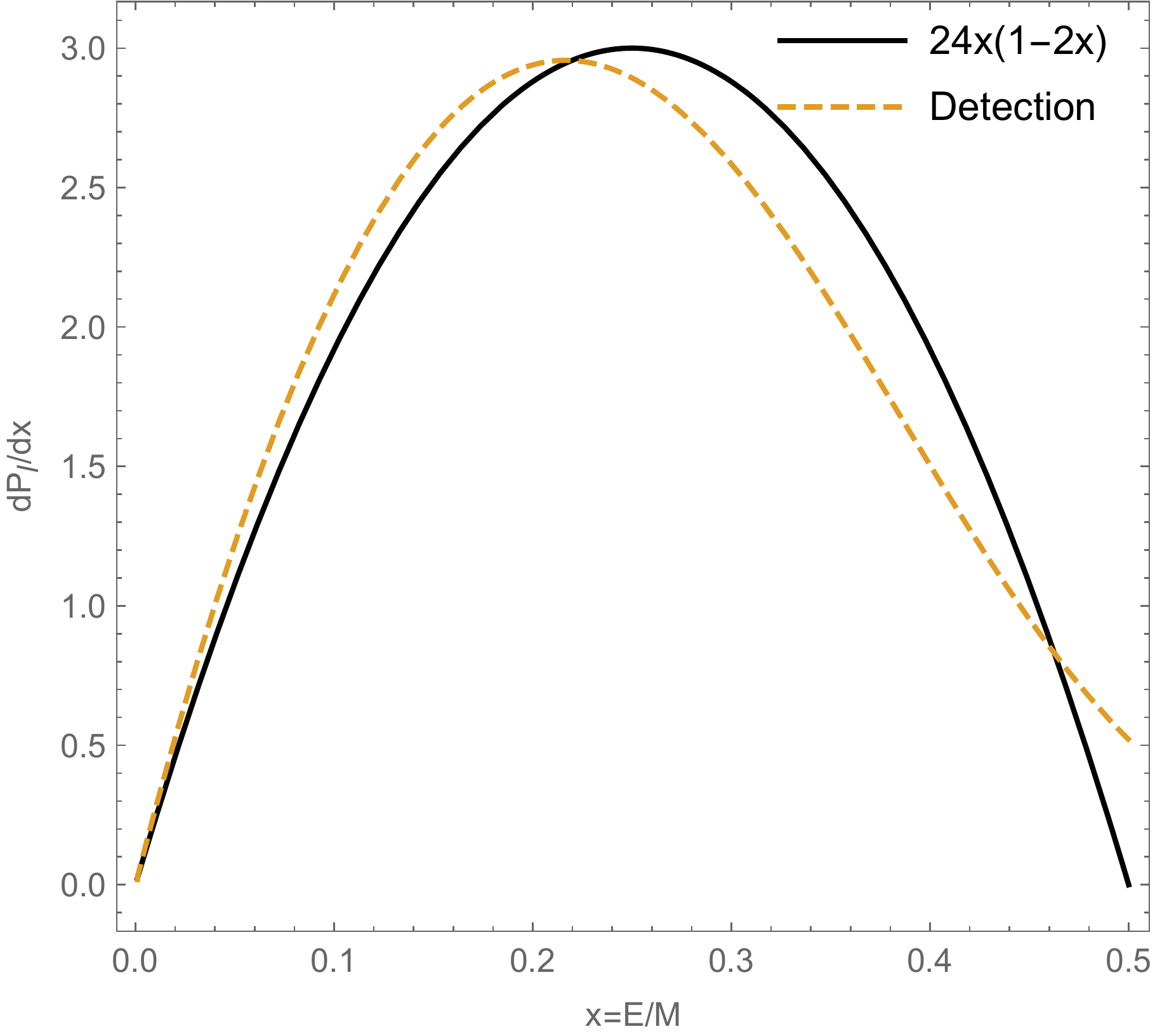}
	\caption{Parabola spectrum.\label{fig:parabola}}
\end{figure}

Although we have illustrated only simple power-law spectra above, we should keep in mind that the general spectra are of polynomial-type, as shown in examples in previous section. We give a further case in which dark matter interacts with standard model particle as $\mathcal{O}_X \bar{\psi}\gamma _\mu \psi A^{\mu}$, where  $\psi$ is SM fermion and $A_\mu$ is the photon field. We can calculate the spectrum of primary photon as 
\begin{equation}
	\frac{dP_l}{dx}=24x(1-2x)\equiv 3\times 8x - 2\times 24x^2,
\end{equation}
where in the last step we have written in the standard form as in Eq.~\ref{eq:xpoly} with $D_1=3$ and $D_2=-2$. The spectrum is shown in Fig.~\ref{fig:parabola} as a parabola (black line). Considering the resolution, the spectrum is displayed as the dashed line.

\section{Summary}\label{sec:summary}
In this paper, we have discussed polynomial features of gamma-ray spectrum from dark matter decay or annihilation. We have found that gamma ray can have general spectral shape as polynomial functions, besides gamma lines, internal bremmstrahlung and box-shaped signals. Our investigation framework is based on kinematic analysis, therefore the results are very generic, model-independent and can be used for a wide class of DM models in which new or mediating degree of freedoms are heavy, compared with standard model particles.

Based the main results, Eq.~\ref{eq:xpoly}, we have shown in Fig.~\ref{fig:spectral} that the polynomial or power-law spectra with an positive index can mimic the line signals in experimental searches for gamma ray or neutrino. This suggests an efficient way for phenomenological studies that we may also start with some polynomial-type gamma-ray spectra for simulation as well as that with a particular DM model or decay/annihilation channel. 




\begin{thebibliography}{10}
	
	\bibitem{Jungman:1995df}
	G.~Jungman, M.~Kamionkowski, and K.~Griest, {\it {Supersymmetric dark matter}},
	\href{http://dx.doi.org/10.1016/0370-1573(95)00058-5}{{\em Phys. Rept.}
		{\bfseries 267} (1996) 195--373}
	[\href{http://arxiv.org/abs/hep-ph/9506380}{{\ttfamily hep-ph/9506380}}].
	
	\bibitem{Bertone:2004pz}
	G.~Bertone, D.~Hooper, and J.~Silk, {\it {Particle dark matter: Evidence,
			candidates and constraints}},
	\href{http://dx.doi.org/10.1016/j.physrep.2004.08.031}{{\em Phys. Rept.}
		{\bfseries 405} (2005) 279--390}
	[\href{http://arxiv.org/abs/hep-ph/0404175}{{\ttfamily hep-ph/0404175}}].
	
	\bibitem{Feng:2010gw}
	J.~L. Feng, {\it {Dark Matter Candidates from Particle Physics and Methods of
			Detection}},
	\href{http://dx.doi.org/10.1146/annurev-astro-082708-101659}{{\em Ann. Rev.
			Astron. Astrophys.} {\bfseries 48} (2010) 495--545}
	[\href{http://arxiv.org/abs/1003.0904}{{\ttfamily arXiv:1003.0904}}].
	
	\bibitem{Bergstrom:2000pn}
	L.~Bergstrom, {\it {Nonbaryonic dark matter: Observational evidence and
			detection methods}},
	\href{http://dx.doi.org/10.1088/0034-4885/63/5/2r3}{{\em Rept. Prog. Phys.}
		{\bfseries 63} (2000) 793}
	[\href{http://arxiv.org/abs/hep-ph/0002126}{{\ttfamily hep-ph/0002126}}].
	
	\bibitem{Hisano:2004ds}
	J.~Hisano, S.~Matsumoto, M.~M. Nojiri, and O.~Saito, {\it {Non-perturbative
			effect on dark matter annihilation and gamma ray signature from galactic
			center}},
	\href{http://dx.doi.org/10.1103/PhysRevD.71.063528}{{\em Phys. Rev.} {\bfseries
			D71} (2005) 063528} [\href{http://arxiv.org/abs/hep-ph/0412403}{{\ttfamily
			hep-ph/0412403}}].
	
	\bibitem{Ishiwata:2008cu}
	K.~Ishiwata, S.~Matsumoto, and T.~Moroi, {\it {High Energy Cosmic Rays from the
			Decay of Gravitino Dark Matter}},
	\href{http://dx.doi.org/10.1103/PhysRevD.78.063505}{{\em Phys. Rev.} {\bfseries
			D78} (2008) 063505} [\href{http://arxiv.org/abs/0805.1133}{{\ttfamily
			arXiv:0805.1133}}].
	
	\bibitem{Atwood:2009ez}
	{\bfseries Fermi-LAT} , W.~B. Atwood {\em et al.}, {\it {The Large Area
			Telescope on the Fermi Gamma-ray Space Telescope Mission}},
	\href{http://dx.doi.org/10.1088/0004-637X/697/2/1071}{{\em Astrophys. J.}
		{\bfseries 697} (2009) 1071--1102}
	[\href{http://arxiv.org/abs/0902.1089}{{\ttfamily arXiv:0902.1089}}].
	
	\bibitem{Bringmann:2012ez}
	T.~Bringmann and C.~Weniger, {\it {Gamma Ray Signals from Dark Matter:
			Concepts, Status and Prospects}},
	\href{http://dx.doi.org/10.1016/j.dark.2012.10.005}{{\em Phys. Dark Univ.}
		{\bfseries 1} (2012) 194--217}
	[\href{http://arxiv.org/abs/1208.5481}{{\ttfamily arXiv:1208.5481}}].
	
	\bibitem{Bergstrom:1997fj}
	L.~Bergstrom, P.~Ullio, and J.~H. Buckley, {\it {Observability of gamma-rays
			from dark matter neutralino annihilations in the Milky Way halo}},
	\href{http://dx.doi.org/10.1016/S0927-6505(98)00015-2}{{\em Astropart. Phys.}
		{\bfseries 9} (1998) 137--162}
	[\href{http://arxiv.org/abs/astro-ph/9712318}{{\ttfamily astro-ph/9712318}}].
	
	\bibitem{Bringmann:2007nk}
	T.~Bringmann, L.~Bergstrom, and J.~Edsjo, {\it {New Gamma-Ray Contributions to
			Supersymmetric Dark Matter Annihilation}},
	\href{http://dx.doi.org/10.1088/1126-6708/2008/01/049}{{\em JHEP} {\bfseries
			01} (2008) 049} [\href{http://arxiv.org/abs/0710.3169}{{\ttfamily
			arXiv:0710.3169}}].
	
	\bibitem{Bringmann:2012vr}
	T.~Bringmann, X.~Huang, A.~Ibarra, S.~Vogl, and C.~Weniger, {\it {Fermi LAT
			Search for Internal Bremsstrahlung Signatures from Dark Matter
			Annihilation}},
	\href{http://dx.doi.org/10.1088/1475-7516/2012/07/054}{{\em JCAP} {\bfseries
			1207} (2012) 054} [\href{http://arxiv.org/abs/1203.1312}{{\ttfamily
			arXiv:1203.1312}}].
	
	\bibitem{Ibarra:2012dw}
	A.~Ibarra, S.~Lopez~Gehler, and M.~Pato, {\it {Dark matter constraints from
			box-shaped gamma-ray features}},
	\href{http://dx.doi.org/10.1088/1475-7516/2012/07/043}{{\em JCAP} {\bfseries
			1207} (2012) 043} [\href{http://arxiv.org/abs/1205.0007}{{\ttfamily
			arXiv:1205.0007}}].
	
	\bibitem{Chu:2012qy}
	X.~Chu, T.~Hambye, T.~Scarna, and M.~H.~G. Tytgat, {\it {What if Dark Matter
			Gamma-Ray Lines come with Gluon Lines?}},
	\href{http://dx.doi.org/10.1103/PhysRevD.86.083521}{{\em Phys. Rev.} {\bfseries
			D86} (2012) 083521} [\href{http://arxiv.org/abs/1206.2279}{{\ttfamily
			arXiv:1206.2279}}].
	
	\bibitem{Ibarra:2013eda}
	A.~Ibarra, H.~M. Lee, S.~López~Gehler, W.-I. Park, and M.~Pato, {\it
		{Gamma-ray boxes from axion-mediated dark matter}},
	\href{http://dx.doi.org/10.1088/1475-7516/2013/05/016}{{\em JCAP} {\bfseries
			1305} (2013) 016} [\href{http://arxiv.org/abs/1303.6632}{{\ttfamily
			arXiv:1303.6632}}].
	
	\bibitem{Toma:2013bka} 
	T.~Toma,
	Phys.\ Rev.\ Lett.\  {\bf 111}, 091301 (2013)
	[arXiv:1307.6181]; A.~Ibarra, T.~Toma, M.~Totzauer and S.~Wild,
	Phys.\ Rev.\ D {\bf 90}, 043526 (2014)
	[arXiv:1405.6917].
	
	\bibitem{Giacchino:2015hvk}
	F.~Giacchino, A.~Ibarra, L.~L. Honorez, M.~H.~G. Tytgat, and S.~Wild,
	{\it {Signatures from Scalar Dark Matter with a Vector-like Quark Mediator}},
	[\href{http://arxiv.org/abs/1511.04452}{{\ttfamily arXiv:1511.04452}}].
	
	\bibitem{Kim:2015usa} 
	D.~Kim and J.~C.~Park,
	[arXiv:1507.07922];
	D.~Kim and J.~C.~Park,
	Phys.\ Lett.\ B {\bf 750}, 552 (2015)
	[arXiv:1508.06640].
	
	\bibitem{Chung:1998ua} 
	D.~J.~H.~Chung, E.~W.~Kolb and A.~Riotto,
	Phys.\ Rev.\ Lett.\  {\bf 81}, 4048 (1998)
	[\href{http://arxiv.org/abs/hep-ph/9805473}{{\ttfamily hep-ph/9805473}}].
	
	\bibitem{Ko:2015nma}
	P.~Ko and Y.~Tang, {\it {IceCube Events from Heavy DM decays through the
			Right-handed Neutrino Portal}},
	\href{http://dx.doi.org/10.1016/j.physletb.2015.10.021}{{\em Phys. Lett.}
		{\bfseries B751} (2015) 81--88}
	[\href{http://arxiv.org/abs/1508.02500}{{\ttfamily arXiv:1508.02500}}].
	
	\bibitem{DMeft}
	M.~Duch, B.~Grzadkowski, and J.~Wudka,
	{\it {Classification of effective operators for interactions between the
			Standard Model and dark matter}},
	[\href{http://arxiv.org/abs/1412.0520}{{\ttfamily arXiv:1412.0520}}].
	
	\bibitem{Aisati:2015ova}
	C.~E. Aisati, M.~Gustafsson, T.~Hambye, and T.~Scarna,
	{\it {Dark Matter Decay to a Photon and a Neutrino: the Double Monochromatic
			Smoking Gun Scenario}},  [\href{http://arxiv.org/abs/1510.05008}{{\ttfamily
			arXiv:1510.05008}}].
	
	\bibitem{Pythia64:2006za}
	T.~Sjostrand, S.~Mrenna, and P.~Z. Skands, {\it {PYTHIA 6.4 Physics and
			Manual}},
	\href{http://dx.doi.org/10.1088/1126-6708/2006/05/026}{{\em JHEP} {\bfseries
			05} (2006) 026} [\href{http://arxiv.org/abs/hep-ph/0603175}{{\ttfamily
			hep-ph/0603175}}].
	
	\bibitem{Ackermann:2015lka}
	{\bfseries Fermi-LAT} , M.~Ackermann {\em et al.}, {\it {Updated search for
			spectral lines from Galactic dark matter interactions with pass 8 data from
			the Fermi Large Area Telescope}},
	\href{http://dx.doi.org/10.1103/PhysRevD.91.122002}{{\em Phys. Rev.} {\bfseries
			D91} no.~12, (2015) 122002}
	[\href{http://arxiv.org/abs/1506.00013}{{\ttfamily arXiv:1506.00013}}].
	
\end{thebibliography}

\providecommand{\href}[2]{#2}\begingroup\raggedright
\endgroup
\end{document}